\documentclass[aps,prb,twocolumn,groupedaddress]{revtex4-1}

\usepackage{amsmath}
\usepackage{graphicx}
\usepackage{color}

\begin{document}

\title{Weak measurement of quantum superposition states in graphene}

\author{Maxim Trushin, Johannes B\"ulte, and Wolfgang Belzig}
\affiliation{University of Konstanz, Fachbereich Physik, M703 D-78457 Konstanz}

\date{\today}

\begin{abstract}
We employ a weak measurement approach to demonstrate the very existence of the photoexcited interband superposition states in intrinsic graphene.
We propose an optical two-beam setup where such measurements are possible and derive an explicit formula for the differential optical absorption that contains a signature of such states.
We provide an interpretation of our results in terms of a non-Markovian weak measurement formalism applied to the pseudospin degree of freedom
coupled with an electromagnetic wave.
\end{abstract}

\maketitle

\section{Introduction}
The past decade has witnessed many experiments addressing coherent quantum superpositions in single quantum systems directly.
Among these are systems with photons \cite{Resch2004,Pryde2005,Kocsis2011,Goette2012,Lee2012}, with nuclear spins \cite{NucSpin1,NucSpin2},
and, more recently, in electronic setups \cite{Shpitalnik:2008jv,Weisz:2013tt,Romito:2014fn}.
Since the quantum system should be left unchanged during the coherent evolution the concept of weak measurements was introduced nearly 30 years ago \cite{Aharonov1988} 
and still remains a highly active topic \cite{RevModPhys.86.307,PhysRevLett.93.056803,RevModPhys.82.1155,Bliokh2013,Ferrie2014}. 
In this paper, we would like to add another possible manifestation of this problem by considering an optical measurement of the photocarriers in semiconductors,
which can be readily implemented. To this end, we propose an experiment to weakly measure the photoexcited interband (electron-hole) superposition states in intrinsic graphene. 
To the best of our knowledge, these electron-hole pairs have never been considered from the weak measurement point of
view, despite the very extensive literature on semiconductor optics. 
By establishing a detailed correspondence between the optical excitation-absorption process and the recently developed non-Markovian weak measurement formalism 
\cite{PhysRevLett.110.250404} we find the signature of the coherent oscillation in the optical absorption signal.

\begin{figure}
\includegraphics[width=\columnwidth]{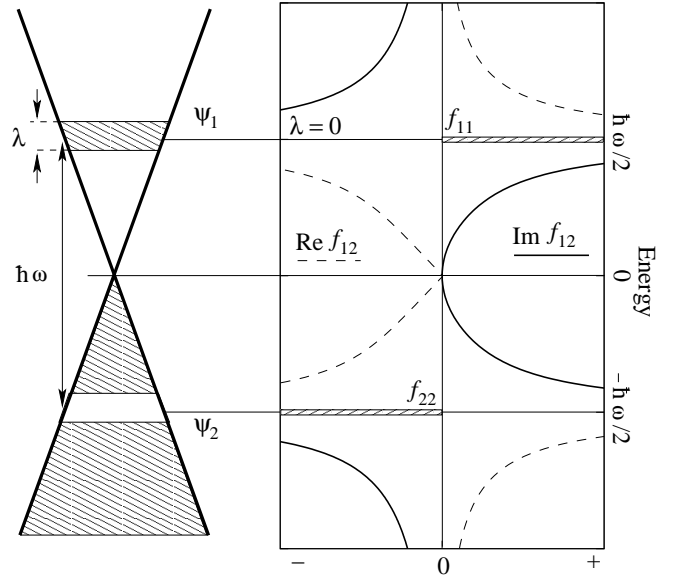}
\caption{\label{Fig1} 
Direct interband excitations result not only in the photocarrier occupations $f_{11}$, $f_{22}=-f_{11}$ at energies $\pm \hbar\omega/2$, but 
also in the coherences $f_{12}$,  $f_{21}=f_{12}^*$ representing the superpositions between conduction- and valence-band eigenstates.
The matrix $f_{ij}$ can be seen as a density matrix written in the eigenstate basis of $\hat H_0$. In the right diagram the density-matrix components are shown
schematically assuming monochromatic radiation ($\lambda=0$, i.e. $f_{ii}$ shrinks to a $\delta$-function).
Note that $\mathrm{Re} f_{12}$ vanishes faster than $\mathrm{Im} f_{12}$ when the energy increases,
i.e. the latter dominates in the high-energy tail of the density matrix representing the signature of the quantum superposition states.
}
\end{figure}
\begin{figure}
\includegraphics[width=\columnwidth]{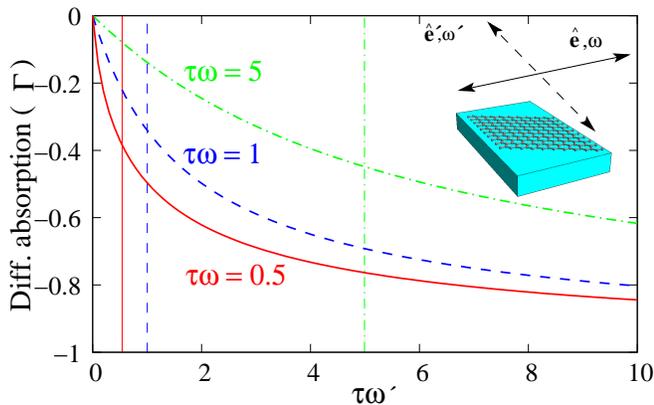}
\caption{\label{Fig2} 
Graphene is irradiated by two cw monochromatic light beams having the frequencies $\omega$ and $\omega'$ and linear polarizations perpendicular to each other.
One of the beams creates an initial photocarrier distribution, whereas another is used as a weak measurement probe.
The curves represent relative optical absorption difference vs. detector's excitation frequency $\omega'$ 
in units of $1/\tau$ (dephasing rate) at zero temperature and zero doping, as given by Eq.~(\ref{A0}).
The differential relative absorption is given in units of $\Gamma$ [see Eq.~(\ref{strength})].
The verticals show the initial photocarrier occupation positions that have shrunk
to zero due to the perfectly monochromatic radiation.
}
\end{figure}

To perform a weak measurement \cite{Tamir2013} of a quantum state $\psi$ the detector must also represent a quantum-mechanical object described by a wave function $\psi_d$.
The measured system interacts with the detector, and it is the detector's wave function $\psi_d$ that is collapsed by a strong measurement.
Due to the interaction between these two states the outcome of the measurement on $\psi_d$ contains some information about coherence of the original quantum state $\psi$.
In our setup, the role of the quantum  state $\psi$ is played by a 2$\times$2 photocarrier density matrix written in the 
eigenstate basis of the massless Dirac Hamiltonian for charge carriers in graphene. 
Once an external radiation source is applied, the matrix represents a sum $f^{(0)}_{ij}+f_{ij}[f^{(0)}]$,
where $f^{(0)}$ is the equilibrium distribution, and $f$ is a non-equilibrium addition on top of $f^{(0)}$. 
The matrix $f$ includes non-zero coherences representing interband superpositions (see Fig.~\ref{Fig1}).
The probe radiation in turn changes the photocarrier density matrix to $f^{(0)}_{ij}+f_{ij}[f^{(0)}]+f'_{ij}[f]$, where $f'$ depends on $f$.
Due to the weak interaction with the probe radiation the photocarrier density matrix is left almost unaffected $f'_{ij}\ll f_{ij}$. 
That is what is known as a linear-response regime. In contrast to the classical linear-response absorption measurement,
we take into account coherences and show that they could be detected in such settings.
We explicitly show that the differential relative absorption of the probe radiation given by Eq.~(\ref{A0}) and shown in Fig.~\ref{Fig2}
contains a response from the initial superposition states encoded in $f_{ij}$, $i\neq j$.
We confirm this result using a non-Markovian weak measurement formalism developed recently,
hence, providing an alternative method for weak measurement description.

\section{Model in depth} 
The effective two-level system shown in Fig.~\ref{Fig1} is described by the two-dimensional massless Dirac-like Hamiltonian \cite{Nature2007geim}
$\hat H_0=\hbar v \hat{\sigma}\cdot \hat{\mathbf{k}}$,
where $\hbar\hat{\mathbf{k}}$ is the two-component momentum operator,
$\hat{\sigma}$ is the pseudospin operator constructed from the Pauli matrices, and $v \approx 10^6$ ms$^{-1}$ is the carrier velocity in graphene.
The corresponding eigenvalues and eigenstates are $E_{1,2}=\pm\hbar v k$ and $\psi_{1,2}=\frac{1}{\sqrt{2}}(1,\pm\mathrm{e}^{i\phi})$
with $\tan\phi=k_y/k_x$. The index $1$ ($2$) stands for the conduction (valence) band (see Fig.~\ref{Fig1}).
The pseudospin is parallel (antiparallel) to momentum while being in a conduction-band (valence-band) eigenstate.
Hence, a certain interband coherent state $\psi=a\psi_1+b\psi_2$ we are trying to measure 
represents, at the same time, a pseudospin superposition state.

In order to create and measure such a superposition state we employ an electromagnetic wave 
$\mathcal{E}(t)= \mathcal{E}_0 \left(\mathrm{e}^{i\omega t} + \mathrm{e}^{-i\omega t}\right)\mathbf{\hat e}$ 
with the frequency $\omega$ and amplitude $\mathcal{E}_0$ linearly polarized in the direction $\mathbf{\hat e}$.
The equilibrium carriers are then perturbed by the field-pseudospin interaction \cite{NJP2012trushin}
$\hat H_\mathrm{int}(t)=\hat h\left(\mathrm{e}^{i\omega t} + \mathrm{e}^{-i\omega t}\right)$, where 
$\hat h= \frac{e v \mathcal{E}_0}{2\omega}\hat{\sigma}\cdot \mathbf{\hat e}$ with $e$ being the electron charge.
We assume normal incidence without momentum transfer from photons to electrons
so that we always superimpose the states $\psi_{1,2}$ with the same momentum.
It is convenient to define the coupling parameter $\eta_\omega=\frac{e v \mathcal{E}_0}{2\omega}$ that entangles the electric field and the pseudospin $\hat\sigma$.
In contrast to conventional semiconductors, where a similar coupling constant can also be defined, 
graphene allows tuning $\eta_\omega$ in a very broad range and in a very predictable manner. 
Indeed, the excitation frequency $\omega$ can be shifted
from the ultraviolet to far infrared regions while keeping the photocarrier velocity constant, hence, changing $\eta_\omega$ by orders of magnitude.
In conventional semiconductors the lowest possible $\omega$ is limited by the band gap, and
the photocarrier velocity is dependent on the excitation frequency reflecting their complicated band structure.
Thus, graphene offers an unprecedented opportunity to study the pseudospin (or interband) coherence by optical means.

\section{Creating a pseudospin superposition state}  
The operator $\hat h$ written in the eigenbasis of $\hat H_0$ is given by the matrix
\begin{equation}
\label{h}
\hat h =\eta_\omega\left(
\begin{array}{ll}
\cos(\phi-\theta) & -i\sin(\phi-\theta)\\
i\sin(\phi-\theta) & -\cos(\phi-\theta) \end{array} \right),
\end{equation}
where $\tan\theta = e_{y}/e_{x}$ describes the polarization plane orientation.
The pseudospin-momentum coupling in graphene results in both off-diagonal and diagonal terms in Eq.~(\ref{h})
responsible for the generation of coherences in the electron density matrix.
This is the crucial ingredient that makes the optical generation and detection of superposition states possible.
It is natural to describe the pseudospin coherence by a $2\times 2$ density matrix $\hat\rho$ the evolution of which
is governed by the Liouville--von Neumann equation
$\partial_t \hat\rho+\frac{i}{\hbar}[\hat H_0,\hat \rho(t)] = \hat g (t)$, where the generation rate is given in
the lowest non-zero order in $h$ by \cite{VaskoBook}
\begin{eqnarray}
\label{gen} \nonumber &&
\hat g (t) = -\frac{1}{\hbar^2}\int\limits_{-\infty}^0 dt' \mathrm{e}^{\lambda t'} \left\{\left(\mathrm{e}^{i\omega t'} + \mathrm{e}^{-i\omega t'}\right)  \right.\\
&&
\left. \times [\hat h,\mathrm{e}^{\frac{i}{\hbar}\hat H_0 t'}[\hat h,\hat \rho(t+t')]\mathrm{e}^{-\frac{i}{\hbar}\hat H_0 t'}] \right\}.
\end{eqnarray}
Here, the adiabaticity parameter $\lambda >0$ has been introduced in order to take into account the finite Lorentzian spectral width of the radiation flow. 
We assume that $\lambda\to 0$ at the end of the day.
This approximation corresponds to the cw operation of the laser.
We now write the density matrix $\hat \rho$ and the corresponding Liouville--von Neumann equation in the eigenstate basis of $\hat H_0$ (helicity basis),
and, at the same time, add relaxation  and  dephasing terms.
The equation then reads
\begin{eqnarray}
\nonumber && \frac{\partial}{\partial t}
\left(\begin{array}{cc}
f_{11} & f_{12} \\
f_{21} & f_{22} \end{array} \right)+\frac{i}{\hbar}
\left(\begin{array}{cc}
0 & f_{12}(E_1-E_2) \\
f_{21}(E_2-E_1) & 0 \end{array} \right) \\
&&
= \left(\begin{array}{cc}
g_{11}[f^0] & g_{12}[f^0] \\
g_{21}[f^0] & g_{22}[f^0] \end{array} \right) - 
\left(\begin{array}{ll}
f_{11}/\tau_0 & f_{12}/\tau \\
f_{21}/\tau & f_{22}/\tau_0 \end{array} \right),
\label{f}
\end{eqnarray}
where the functions $f_{ij}$ represent the change in the density matrix $\hat\rho-\hat\rho^{\text{eq}}$ written in the basis of the unperturbed eigenstates of $\hat H_0$, 
$f^0$ is the equilibrium distribution matrix diagonal in the same basis, $\tau_0$ is the relaxation time, and $\tau$ is the dephasing time.
The generation rate is transformed within the rotating wave approximation \cite{Wu2007} and reads 
\begin{eqnarray}
\label{g11}
\nonumber g_{11}[f^0] =  && \frac{2 h_{12} h_{21}}{\hbar^2}\left(f^{(0)}_{22}-f^{(0)}_{11}\right) \\
 \times && \sum\limits_\pm \frac{\lambda}{\lambda^2 + (\Omega \pm \omega)^2}, \\
 \label{g12}
\nonumber g_{12}[f^0]  = && \frac{h_{11} h_{12} + h_{21} h_{22}}{\hbar^2}\left(f^{(0)}_{11}-f^{(0)}_{22}\right) \\
\times && \sum\limits_\pm \frac{1}{\lambda +  i(\Omega \pm \omega)},
\end{eqnarray}
and $g_{22} =-g_{11} $,  $g_{21}=g_{12}^*$.
Here, $\Omega=(E_1 - E_2)/\hbar=2vk$.
Note that $g_{11}$ is proportional to a $\delta$-function at $\lambda \to 0$, whereas $g_{12}$ is not.
Since the interaction matrix $h_{ij}$ contains {\em both} diagonal and non-diagonal terms,
the generation rate $g_{ij}$ does the same.
This is the reason why the light-carrier interaction creates the interband coherent states. 
If the generation $g_{ij}$ were turned off at $t=0$, then the coherences in $f_{ij}$ would be given by 
$f_{12}(t)=f_{12}(0)\exp\left(-t/\tau - i\Omega t \right)$, $f_{21}=f_{12}^*$
representing a rapidly oscillating function with the amplitude determined by the initial condition at $t=0$.
Measuring such an oscillating function could be a challenging task and, therefore, we focus on a steady-state limit,
when $\partial f_{ij}/ \partial t = 0$. In this limit, the solution of Eq. (\ref{f}) is given by 
$f_{12}=\tau g_{12}[f^0]/(1+i\tau\Omega)$, $f_{11}=\tau_0 g_{11}[f^0]$,
$f_{21}=f_{12}^*$, $f_{22}=-f_{11}$ (see Appendix \ref{app_coherences}).
At $\lambda\to 0$, the diagonal terms $f_{ii}$ represent occupations of the corresponding bands
given by the delta function $\delta(\Omega-\omega)$ whereas the coherences 
$f_{ij}$ describe the pseudospin superposition states and remain finite even at $\Omega \neq \omega$.
The explicit expressions for $f_{ij}$ as derived in Appendix \ref{app_coherences} are given by
\begin{equation}
 f_{11}=\frac{2\pi \tau_0}{\hbar^2} h_{12} h_{21}\left(f^{(0)}_{22}-f^{(0)}_{11}\right) \sum\limits_\pm  \delta_\lambda (\Omega \pm \omega),
\end{equation}
and
\begin{equation}
 f_{12}=\frac{h_{11} h_{12} + h_{21} h_{22}}{\hbar^2}\frac{\tau}{1 + i\tau \Omega}\sum\limits_\pm
 \frac{f^{(0)}_{11}-f^{(0)}_{22}}{\lambda +  i(\Omega \pm \omega)}.
\end{equation}

\section{Measuring the superposition state}
The interband coherences schematically depicted in Fig.~\ref{Fig1} are generated by $\hat H_\mathrm{int}$,
which couples the electric field and pseudospin --- the quantity we are trying to measure. The weak measurement is performed by means of the interaction
between the probe electromagnetic wave and pseudospin:
$\hat H'_\mathrm{int}= \hat h'\left(\mathrm{e}^{i\omega' t} + \mathrm{e}^{-i\omega' t}\right)$ 
with $\hat h'=\frac{e v \mathcal{E}'_0}{2\omega'}\hat{\sigma}\cdot \mathbf{\hat e}'$, where $\mathbf{\hat e}'\perp \mathbf{\hat e}$.
The orthogonality between $\mathbf{\hat e}'$ and $\mathbf{\hat e}$ allows us to filter out the absorption of any of these two electromagnetic waves easily.
Similar to $\hat h$, the electric-field amplitude $\mathcal{E}'_0$ and pseudospin $\hat\sigma$ are coupled by the coupling parameter $\eta_{\omega'}=\frac{e v \mathcal{E}'_0}{2\omega'}$.
The probe radiation is assumed to have the intensity $I'_0=c\mathcal{E'}^2_0/8\pi$ much lower than $I_0=c\mathcal{E}^2_0/8\pi$ so that 
the probe almost does not change the coherent state created initially by $\hat h$.
It is also crucial for the probe interaction $\hat h'$ to be weak enough to justify the weak measurement criterion
within the whole frequency interval we are interested in, i.e. the probe intensity should be tuned while changing the probe frequency $\omega'$ so that $\eta_{\omega'}$ remains constant. 
The complete quantum-mechanical weak measurement procedure also requires averaging over many single weak measurements \cite{Tamir2013}.
Since we employ the density matrix (not a wave function) for the quantum state description, the statistics is taken into account automatically within our model.
In the experiment, the cw operating probe serves for the desired statistics.

The probe generation rate $\hat g'(t)$ is formally given by the equation similar to Eq.~(\ref{gen}), but the resulting $g'_{ij}[f]$ 
depends not only on the band occupations $f_{11}$, $f_{22}$ but also on the coherences $f_{12}$, $f_{21}$ introduced above.
The coherences vanish in equilibrium; however, once the steady-state non-equilibrium is created,
the coherences can be observed by measuring the optical absorption calculated from $g'_{ij}[f]$.
The relative optical absorption is defined as $A=I'/I'_0$, where $I'$ is the absorbed intensity per spin/valley channel given by $I'=\hbar \omega' \int d^2 k  g'_{11}/(2\pi)^2$. 
Here, we need only one component of the generation matrix $g'_{ij}$ representing the valence-to-conduction-band transition rate given by 
\begin{eqnarray}
 \label{g11main} &&
g'_{11}[f]=\sum\limits_\pm\frac{h'_{12} h'_{21}}{\hbar^2}\frac{2\lambda\left(f_{22} - f_{11}\right)}{\lambda^2 + ( \Omega \pm \omega')^2}\\
\nonumber &&
-2i\sum\limits_\pm\frac{h'_{12}h'_{22}}{\hbar^2}\frac{(\Omega \pm \omega') \mathrm{Re} f_{12}   - \lambda \mathrm{Im} f_{12}}{\lambda^2 + (\Omega \pm \omega')^2}.
\end{eqnarray}
The first term formally looks similar to Eq.~(\ref{g11}) but the equilibrium occupations $f^0_{ii}$ are substituted by their steady-state counterparts $f_{ii}$
determined by the relaxation time $\tau_0$.
This term dominates in the absorption  at $\omega = \omega'$ and corresponds to the conventional signal in the optical pump-probe measurements \cite{Trushin2015}.
The second term contains quantum information about the initial steady state via coherences $f_{ij}$ determined by the dephasing time $\tau$.
At $\lambda\to 0$, $\mathrm{Im} f_{12}$ and $\mathrm{Re} f_{12}$ are weighted in the relative absorption by the $\delta$-function $\delta(\Omega\pm \omega')$ 
and the principle value $1/(\Omega\pm\omega')$, respectively. 
$\mathrm{Im} f_{12}$ dominates in the absorption at $\omega' >> \omega$.

The description in terms of  a non-Markovian weak measurement formalism \cite{PhysRevLett.110.250404} can provide insight into the interaction of the probe beam and the pseudospin. 
Starting from a microscopic model \cite{unpublished1} (see Appendix \ref{app_measurement}), 
the field-field correlator of the outgoing electric field after the interaction with the graphene is given by 
\begin{eqnarray}
\label{correlator1} &&
 \langle \mathcal{E}'(0)\mathcal{E}'(0) \rangle_w \\
 \nonumber && 
 = \frac{1}{2\hbar^2}\int dt \int dt' \chi(0,t) \chi(0,t') \langle \{\hat\sigma_x(t),\hat\sigma_x(t')\}\rangle  \\
 \nonumber && -\frac{2i}{\hbar^2}\int dt \int dt' \theta(t-t') \chi(0,t)S(0,t') \langle [\hat\sigma_x(t),\hat\sigma_x(t')] \rangle,
\end{eqnarray}
where $\chi(0,t)=-i \frac{\eta_{\omega'}}{\sqrt{N}}\theta(-t)\mathrm{e}^{\lambda t}\langle [\mathcal{\hat E}'(0),\mathcal{\hat E}'(t)] \rangle$ is the response function and
$S(0,t)=\frac{\eta_{\omega'}}{2\sqrt{N}}\mathrm{e}^{\lambda |t|}\langle \{ \mathcal{\hat E}' (0), \mathcal{\hat E}'(t) \} \rangle$ is the noise function of the probe beam.
Note, that the dimensionless electric field $\mathcal{\hat E}'=\frac{i}{\sqrt{2}}(\hat a \mathrm{e}^{-i\omega' t}-\hat a^\dagger \mathrm{e}^{i\omega' t})$
is quantized in this formalism and the probe intensity is determined by the photon number $N=\langle \hat a^\dagger \hat a \rangle +\frac{1}{2}$ 
to which we normalize here. The parameter $\lambda$ takes into account a Lorentzian broadening of the laser beam.
The square and curly brackets denote commutator and anticommutator, respectively.
The first term in Eq.~(\ref{correlator1}) does not contribute to the differential absorption
because the trace is taken over the steady-state distribution function $f_{ij}$ determined by Eq.~(\ref{f}) resulting in $\mathrm{Tr}\,f=0$.
Hence, we focus on the second term. It describes the self-interaction of the electric field mediated by the graphene 
and therefore contains the pseudospin response function $\theta(t-t')\langle [\hat\sigma_x(t),\hat\sigma_x(t')] \rangle$ which is sensitive to the presence of coherence.
Therefore it adds a signature of the quantum state of the graphene to the outgoing electric field.
The steady-state approximation is realized as $\langle [\hat\sigma_x(t),\hat\sigma_x(t')] \rangle \to \langle [\hat\sigma_x(t-t'),\hat\sigma_x(0)] \rangle$.
The pseudospin is considered in the eigen state basis of $\hat H_0$ and its 
dynamics is governed by the equation $d_t \hat\sigma_i (t)  = \frac{i}{\hbar}[\hat H_0, \hat \sigma_i]$.
The statistical average is performed as $\langle \sigma_i(t) \rangle =\mathrm{Tr}\,\hat \sigma_i(t) f $.
In particular, $\langle \sigma_x \rangle=(f_{11}-f_{22})\cos\phi  - 2 \sin\phi\, \mathrm{Im}f_{12}$,
$\langle \sigma_y \rangle=(f_{11}-f_{22}) \sin\phi  + 2 \cos\phi\, \mathrm{Im}f_{12}$,
and $\langle \sigma_z \rangle = 2 \mathrm{Re} f_{12}$.
The resulting correlator reads
\begin{eqnarray}
\nonumber && \langle \mathcal{E}'(0) \mathcal{E}'(0) \rangle_w  = \frac{\eta_{\omega'}^2 }{\hbar^2}\int\limits_{-\infty}^0 dt \int\limits_{-\infty}^t dt' \sin(\omega't) \cos(\omega't')\\
\nonumber && \times  4\mathrm{e}^{\lambda t}\mathrm{e}^{\lambda t'} \left( \langle \sigma_y \rangle \sin\phi \sin[\Omega(t-t')]\right. \\
&& \left. +  \langle \sigma_z \rangle \cos\phi \sin\phi  \{\cos[\Omega(t-t')] - 1\} \right).
\label{correlator2}
\end{eqnarray}
We divide the correlator by the laser pulse duration $\frac{1}{\lambda}$.
Then, we employ the rotating wave approximation when integrating Eq.~(\ref{correlator2}). The result for small $\lambda\ll\omega,\omega'$ reads
\begin{eqnarray}
\nonumber  \langle \mathcal{E}'(0) \mathcal{E}'(0) \rangle_w & = & 
 -\pi \frac{\eta_{\omega'}^2}{2\hbar^2}\langle \sigma_y \rangle \sin\phi \sum\limits_\pm \delta(\omega'\pm \Omega)  \\
 && +\frac{\eta_{\omega'}^2}{\hbar^2} \langle \sigma_z \rangle  \frac{\cos\phi \sin\phi \,\Omega^2}{\omega'(\omega'^2 - \Omega^2)},
\label{alternate}
\end{eqnarray}
cf. Eq.~(\ref{g11main}). 
In the simplest case of a zero-temperature equilibrium state, we have  $f_{22}=1$, $f_{11}=f_{12}=f_{21}=0$, hence,
$\langle \sigma_y \rangle^{\text{eq}}=-\sin\phi $, $\langle \sigma_z \rangle^{\text{eq}} = 0$.
To calculate the total absorbed energy per unit time and unit area from Eq.~(\ref{alternate}) we integrate over all momenta and multiply by $\hbar\omega'$.

For isotropic $\langle\sigma_z\rangle$ or $\langle\sigma_y\rangle$ the corresponding parts average out by this momenta integration. However, our excitations depend on the direction [c.f. $\langle \sigma_z \rangle\propto h_{11}h_{12}+h_{21}h_{22}=-2i\cos(\phi-\theta)\sin(\phi-\theta)$] and the occurring integrals in Eq. (\ref{alternate}), $\int_0^{2\pi}\text{d}\phi\sin(\phi-\theta)\cos(\phi-\theta)\sin(\phi)\cos(\phi)=\frac{\pi}{4}\cos(2\theta)$ and $\int_0^{2\pi}\text{d}\phi\sin^2(\phi-\theta)\sin^2(\phi)=\frac{\pi}{4}(2-\cos(2\theta))$, do not vanish for orthogonal pump and probe beams $\theta=\frac{\pi}{2}$.

To obtain the relative absorption we divide the result by the incident intensity $I_0'$ and multiply by $4$
for the valley and spin degeneracy. The equilibrium absorption then reads
\begin{equation}
 A_0=4 \frac{\hbar\omega'}{I_0'}\int\frac{\text{d}^2k}{4\pi^2}\langle\mathcal{E}'(0)\mathcal{E}'(0)\rangle_w^{\text{eq}} = \frac{\pi e^2}{\hbar c},
\end{equation}
which is a known result \cite{Science2008nair}. Most saliently, either of Eqs. (\ref{alternate}) and (\ref{g11main}) contains non-equilibrium quantum information
encoded in the coherences $f_{12}$ and $f_{21}$ the manifestation of which can also be seen in the absorption.

 We focus on the relative absorption difference $(A-A_0)/A_0$ at $\lambda=0$ and $\omega\neq \omega'$
when the first term in Eq.~(\ref{g11main}) vanishes so that the occupations do not contribute.
To neglect the influence of occupations at $\lambda \neq 0$ the probe frequency $\omega'$ must be fixed further away from $\omega$.
The differential absorption is negative due to the so-called Pauli bleaching and reads 
\begin{equation}
 \label{A0}
 \frac{A-A_0}{\Gamma\, A_0} =  - 
\frac{\omega' \tau^2}{\omega'^2-\omega^2}\left(\frac{\omega'^3}{1+\tau^2\omega'^2} - \frac{\omega^3}{1+\tau^2\omega^2} \right) 
$$
$$
- \frac{2\omega' \tau }{\pi}
 \frac{\omega'^2 \ln (\tau\omega') - \omega^2 \ln (\tau\omega) + \tau^2 \omega'^2 \omega^2 \ln(\omega'/\omega)}
 {(1+\tau^2\omega^2)(1+\tau^2\omega'^2)(\omega^2-\omega'^2)},
\end{equation} 
where $A_0=\pi e^2/(\hbar c)$ is the equilibrium absorption, and
\begin{equation}
 \label{strength}
 \Gamma = \frac{1}{A_0 I'_0} \frac{\eta_{\omega'}^2 \eta_{\omega}^2}{\hbar^3 v^2}
\end{equation}
is a constant that controls the overall signal strength via the coupling parameters $\eta_{\omega}$, $\eta_{\omega'}$.
We estimate $\Gamma$ to be of the order of $10^{-5}$ for typical cw lasers with the power of about 1 mW
focused on a $\mu$m-size spot and the frequencies $\omega$, $\omega'$ being of the order of 0.1 and 1 eV, respectively.
Measuring such a low signal is a challenging task;
however, the  relative differential transmission of the order of $10^{-4}$ has already been measured recently 
on graphene using an optical pump-probe setup \cite{Trushin2015}. Note that $\Gamma$ can be changed in a very broad
range by playing with $\eta_{\omega}$ and $\eta_{\omega'}$, as there is no theoretical limit for excitation frequencies from below (no band gap).

\section{Discussion}
Equation (\ref{A0}) is among the main results of our paper. 
This signal is a response to the pseudospin superposition states created by $\eta_{\omega}$ and measured by $\eta_{\omega'}$.
The signal should vanish if such states were not there.
The relative absorption (\ref{A0}) is solely due to the pseudospin coherence
because the conventional occupation-related dip shrinks to zero at $\lambda=0$. 
For this reason Eq.~(\ref{A0}) does not depend on $\tau_0$ and vanishes at $\tau\to 0$ (very fast dephasing).
On the other hand, the signal saturates at $\tau\to\infty$ and equals to
\begin{equation}
\label{saturation}
\frac{A-A_0}{A_0} = -\Gamma \frac{\omega'}{\omega + \omega'}.
\end{equation}
The value of the dephasing time $\tau$ is unknown but it enters Eq.~(\ref{A0}) only in products with $\omega$ and $\omega'$
resulting in the universal behavior shown in Fig.~\ref{Fig2}. One can see that reducing $\omega'$ 
affects the signal in the same way as faster dephasing does. On the other hand, faster dephasing 
is equivalent to higher $\omega$. This effect can also be seen in Eq.~(\ref{saturation})
vanishing at $\omega\to\infty$ even though dephasing is absent there.

Equation (\ref{A0}) can be further simplified at $\omega' \gg \omega$ as
\begin{equation}
 \label{simpler}
 \frac{A-A_0}{A_0} =  -\Gamma  \frac{(\tau \omega')^2}{1+(\tau\omega')^2} \left[1- \frac{2 \ln (\tau\omega')}{\pi \tau\omega'} \right].
\end{equation}
As it is expected from Eq.~(\ref{saturation}), the signal (\ref{simpler}) saturates at $\omega'\tau \to \infty$ as $(A-A_0)/A_0 =  -\Gamma$.
Let us recall that in order to keep $\eta_{\omega'}$ constant in the formal limit $\omega'\tau \to \infty$ we should increase $I'_0$ to infinity as well
making this regime experimentally irrelevant. The regime  $\omega' \gg \omega$ is most convenient for the pseudospin coherence observation, where 
the differential absorption is approaching its maximum and, most importantly, the 
conventional occupation dip at $\omega' = \omega$ is not present even if it is broadened by relaxation.


\section{Conclusion}  To conclude, we have demonstrated a framework for a weak measurement description, 
where the quantum-mechanical coherence and carrier statistics are treated using the Liouville--von Neumann equation
and, alternatively, the field-field correlator. We apply this approach to intrinsic graphene
the peculiar band structure of which makes it possible to generate and observe the quantum-mechanical pseudospin coherence. 
The weak measurement of pseudospin coherence is provided by coupling the pseudospin to the electric field of an electromagnetic wave.
This coupling is what makes graphene special and, in combination with the results presented above,
provides a reliable solid-state platform for fundamental quantum research.

\acknowledgements
We thank Daniele Brida for multiple discussions of these results. This paper was supported financially by the Center of Applied Photonics, 
the European Research Council Advanced Grant UltraPhase of Alfred Leitenstorfer, and the
Deutsche Forschungsgemeinschaft through SFB 767.

\appendix 

\begin{widetext}

\section{Optically excited coherences}
\label{app_coherences}

Equation (2) in the main text and the Liouville--von Neumann equation itself are the operator equations, and to get the corresponding equation for the distribution function
we rewrite the both in the eigenbasis of $\hat H_0$. The density matrix $\hat\rho$ is then transformed as
$$
\hat \rho\to
\left(
\begin{array}{ll}
f_{11} & f_{12}\\
f_{21} & f_{22}
\end{array}\right).
$$
The terms in the left-hand side of the Liouville--von Neumann equation are transformed as
$$
\frac{\partial \hat\rho}{\partial t} \to \left(
\begin{array}{ll}
\partial_t f_{11} & \partial_t f_{12}\\
\partial_t f_{21} & \partial_t f_{22}
\end{array}\right),
$$
$$
\frac{i}{\hbar}[\hat H_0,\hat \rho(t)]\to \frac{i}{\hbar}\left(
\begin{array}{ll}
0 & f_{12}(E_1-E_2)\\
f_{21}(E_2 - E_1) & 0
\end{array}\right).
$$
Here, $E_{1,2}=\pm\hbar v k$ are conduction- and valence-band dispersions.
Note, that the integrand in the generation rate consists of two similar terms $\propto \mathrm{e}^{\pm i\omega t'}$. Each of these two terms 
consists in turn of four terms transformed as
\begin{equation}
\label{1}
\hat h\mathrm{e}^{\frac{i}{\hbar}\hat H_0 t'}\hat h\hat \rho\mathrm{e}^{-\frac{i}{\hbar}\hat H_0 t'}\to
\left( \begin{array}{ll}
(f_{11} h_{12} h_{21} +  f_{21} h_{12} h_{22})\mathrm{e}^{\frac{it'}{\hbar}(E_2-E_1)} & (f_{12} h_{11} h_{11} + f_{22} h_{11} h_{12})\mathrm{e}^{\frac{it'}{\hbar}(E_1-E_2)}  \\
(f_{11} h_{22} h_{21}  + f_{21} h_{22} h_{22})\mathrm{e}^{\frac{it'}{\hbar}(E_2-E_1)}  & (f_{12} h_{21} h_{11}  + f_{22} h_{21} h_{12} )\mathrm{e}^{\frac{it'}{\hbar}(E_1-E_2)}
\end{array}\right),
\end{equation}
\begin{equation}
\label{2}
-\hat h\mathrm{e}^{\frac{i}{\hbar}\hat H_0 t'}\hat\rho  \hat h \mathrm{e}^{-\frac{i}{\hbar}\hat H_0 t'}\to
-\left( \begin{array}{ll}
(f_{21} h_{12} h_{11} + f_{22} h_{12} h_{21})\mathrm{e}^{\frac{it'}{\hbar}(E_2-E_1)} & (f_{11} h_{11} h_{12} + f_{12} h_{11} h_{22})\mathrm{e}^{\frac{it'}{\hbar}(E_1-E_2)}  \\
(f_{21} h_{22} h_{11} + f_{22} h_{22} h_{21})\mathrm{e}^{\frac{it'}{\hbar}(E_2-E_1)}  & (f_{11} h_{21} h_{12} + f_{12} h_{21} h_{22})\mathrm{e}^{\frac{it'}{\hbar}(E_1-E_2)}
\end{array}\right),
\end{equation}
\begin{equation}
\label{3}
-\mathrm{e}^{\frac{i}{\hbar}\hat H_0 t'} \hat h \hat \rho   \mathrm{e}^{-\frac{i}{\hbar}\hat H_0 t'} \hat h\to
-\left( \begin{array}{ll}
(f_{12} h_{11} h_{21} + f_{22} h_{12} h_{21})\mathrm{e}^{\frac{it'}{\hbar}(E_1-E_2)} & (f_{12} h_{11} h_{22} + f_{22} h_{12} h_{22})\mathrm{e}^{\frac{it'}{\hbar}(E_1-E_2)}  \\
(f_{11} h_{21} h_{11} + f_{21} h_{22} h_{11})\mathrm{e}^{\frac{it'}{\hbar}(E_2-E_1)}  & (f_{11} h_{21} h_{12} + f_{21} h_{22} h_{12})\mathrm{e}^{\frac{it'}{\hbar}(E_2-E_1)}
\end{array}\right),
\end{equation}
\begin{equation}
\label{4}
\mathrm{e}^{\frac{i}{\hbar}\hat H_0 t'}  \hat \rho \hat h  \mathrm{e}^{-\frac{i}{\hbar}\hat H_0 t'}\hat h\to
\left( \begin{array}{ll}
(f_{11} h_{12} h_{21} + f_{12} h_{22} h_{21})\mathrm{e}^{\frac{it'}{\hbar}(E_1-E_2)} & (f_{11} h_{12} h_{22} + f_{12} h_{22} h_{22})\mathrm{e}^{\frac{it'}{\hbar}(E_1-E_2)}  \\
(f_{21} h_{11} h_{11} + f_{22} h_{21} h_{11})\mathrm{e}^{\frac{it'}{\hbar}(E_2-E_1)}  & (f_{21} h_{11} h_{12} + f_{22} h_{21} h_{12})\mathrm{e}^{\frac{it'}{\hbar}(E_2-E_1)}
\end{array}\right).
\end{equation}
Here, the matrix $h$ is given by Eq. (1).

Once we are in equilibrium $f_{11}=f^{(0)}_{11}$,  $f_{22}= f^{(0)}_{22}$, $f_{12}= f_{21}=0$, where
$f^{(0)}_{11}=f^{(0)}(E_1)$, $f^{(0)}_{22}=f^{(0)}(E_2)$ are the Fermi-Dirac distributions.
Thus, the $t'$ dependence remains only in the exponents, and the excitation rate reads
\begin{equation}
\label{g11}
 g_{11}= \frac{2h_{12} h_{21}}{\hbar^2}\left(f^{(0)}_{22}-f^{(0)}_{11}\right)\sum\limits_\pm \frac{\lambda}{\lambda^2 + \frac{1}{\hbar^2}(\Delta E \pm E_\omega)^2}
 \to \frac{2\pi}{\hbar}h_{12} h_{21}\left(f^{(0)}_{22}-f^{(0)}_{11}\right)\delta_\lambda (\Delta E \pm E_\omega ).
\end{equation}
\begin{equation}
\label{g_22}
 g_{22}= \frac{2h_{21} h_{12}}{\hbar^2}\left(f^{(0)}_{11}-f^{(0)}_{22}\right)\sum\limits_\pm \frac{\lambda}{\lambda^2 + \frac{1}{\hbar^2}( \Delta E \pm E_\omega)^2}
 \to \frac{2\pi}{\hbar}h_{21} h_{12}\left(f^{(0)}_{11}-f^{(0)}_{22}\right)\delta_\lambda (\Delta E \pm E_\omega).
\end{equation}
Here,
\begin{equation}
 \delta_\lambda (\Delta E \pm E_\omega) = \lim\limits_{\lambda\to 0} \frac{1}{\pi}\frac{\lambda}{\lambda^2 + \frac{1}{\hbar^2}(\Delta E \pm E_\omega)^2},
\end{equation}
where $\Delta E=E_1 - E_2>0$. Note, that $g_{22}=-g_{11}$ and since $E_\omega=\hbar \omega >0$ some $\delta$-functions do not contribute at $\lambda=0$.
Further on, the coherence generation rates read
\begin{equation}
\label{g_12}
 g_{12}= \frac{h_{11} h_{12} + h_{21} h_{22}}{\hbar^2}\left(f^{(0)}_{11}-f^{(0)}_{22}\right)
 \sum\limits_\pm \frac{1}{\lambda +  \frac{i}{\hbar}(\Delta E\pm E_\omega)},
\end{equation}
\begin{equation}
\label{g_21}
 g_{21}= \frac{h_{22} h_{21} + h_{12} h_{11}}{\hbar^2}\left(f^{(0)}_{22}-f^{(0)}_{11}\right)\sum\limits_\pm  \frac{1}{\lambda -  \frac{i}{\hbar}(\Delta E \pm E_\omega)}.
\end{equation}
Note, that $g_{21}=g_{12}^*$.
The distributions $f_{ij}$ can be found from
\begin{equation}
 -\frac{1}{\tau_0}f_{ij} + g_{ij}=0, \quad i=j,
\end{equation}
\begin{equation}
 \frac{i}{\hbar} f_{ij}(E_i - E_j) = -\frac{1}{\tau}f_{ij} + g_{ij}, \quad
 i\neq j,
\end{equation}
where $\tau$ is the pseudospin relaxation time.
Thus,
\begin{equation}
\label{f11}
 f_{11}=\frac{2\pi \tau_0}{\hbar} h_{12} h_{21}\left(f^{(0)}_{22}-f^{(0)}_{11}\right) \sum\limits_\pm  \delta_\lambda (\Delta E \pm  E_\omega ).
\end{equation}
\begin{equation}
\label{f22}
 f_{22}=\frac{2\pi \tau_0}{\hbar} h_{21} h_{12}\left(f^{(0)}_{11}-f^{(0)}_{22}\right)\sum\limits_\pm  \delta_\lambda (\Delta E \pm  E_\omega ),
\end{equation}
\begin{equation}
 \label{f12}
 f_{12}=\frac{h_{11} h_{12} + h_{21} h_{22}}{\hbar^2}\frac{\tau}{1 + \frac{i}{\hbar}\tau \Delta E}\sum\limits_\pm
 \frac{f^{(0)}_{11}-f^{(0)}_{22}}{\lambda +  \frac{i}{\hbar}(\Delta E \pm E_\omega)},
\end{equation}
\begin{equation}
 \label{f21}
 f_{21}=\frac{h_{22} h_{21} + h_{12} h_{11}}{\hbar^2}\frac{\tau}{1 - \frac{i}{\hbar}\tau \Delta E} 
 \sum\limits_\pm \frac{f^{(0)}_{22}-f^{(0)}_{11}}{\lambda -  \frac{i}{\hbar}(\Delta E \pm E_\omega)},
\end{equation}
where the general relations $f_{22}=-f_{11}$, $f_{21}=f_{12}^*$ hold.

\section{Non-Markovian weak measurement}
\label{app_measurement}

We derive the cross-correlation of a non-Markovian weak measurement from a microscopic model. 
In our paper the detector is given by the electric field of a laser beam $\mathcal{\hat E}$ which is coupled to the pseudospin $\hat\sigma$ of a graphene system 
and subsequently measured by a photodetector. In order to model the cross-correlation we label two electric fields (two detectors) with $a$ and $b$ so that the interaction Hamiltonian can be written as
\begin{equation}
\hat H_{\text{int}}=\eta_a \mathcal{\hat E}_a \hat\sigma+\eta_b\mathcal{\hat E}_b\hat\sigma,
\end{equation}
where the $\eta_i$ denote the coupling strength. The probability to find the outcomes $\mathcal{E}_a$ and $\mathcal{E}_b$ is given by
\begin{equation}
p_{\mathcal{E}_a\mathcal{E}_b}(t)=\text{Tr}\left\{\hat K_{\mathcal{E}_a}\hat K_{\mathcal{E}_b} \hat U(t)\hat\rho_0\hat\rho_a\hat\rho_b\hat U^{\dagger}(t)\hat K_{\mathcal{E}_b}^{\dagger}\hat K_{\mathcal{E}_a}^{\dagger}\right\},
\end{equation}
with $\hat U(t)=\hat U_0(t)\hat U_I(t)$, where $\hat U_0$ is the unitary time evolution of the undisturbed systems and $\hat U_I=\mathcal{T} e^{-i\int^{t}\text{d}t'\, H_{\text{int}}^{I}(t')/\hbar}$ labels the interaction with the time-ordering $\mathcal{T}$. The readout of the electric field after the interaction with the pseudospin is denoted by the projection operators $\hat\Pi_{\mathcal{E}_i}=|\mathcal{E}\rangle\langle\mathcal{E}|$. Following the weak coupling assumption we expand the time evolution in the interaction picture to second order in $\eta$
\begin{align}
p_{\mathcal{E}_a\mathcal{E}_b}=&-\frac{\eta_a\eta_b}{\hbar^2}\int^{t}\text{d}t'\int^{t}\text{d}s'\,\langle[\hat \Pi_{\mathcal{E}_a}(t),\mathcal{\hat E}_a(t')]\rangle\langle[\hat \Pi_{\mathcal{E}_b}(t),\mathcal{\hat E}_b(s')]\rangle\langle\{\hat\sigma(t'),\hat\sigma(s')\}\rangle/2\nonumber\\
&-\frac{\eta_a\eta_b}{\hbar^2}\int^{t}\text{d}t'\int^{t}\text{d}s'\,\theta(t'-s')\langle\{\hat\Pi_{\mathcal{E}_a}(t),\mathcal{\hat E}_a(t')\}\rangle\langle[\hat \Pi_{\mathcal{E}_b}(t),\mathcal{\hat E}_b(s')]\rangle\langle[\hat\sigma(t'),\hat\sigma(s')]\rangle/2\nonumber\\
&-\frac{\eta_a\eta_b}{\hbar^2}\int^{t}\text{d}t'\int^{t}\text{d}s'\,\theta(t'-s')\langle[\hat\Pi_{\mathcal{E}_a}(t),\mathcal{\hat E}_a(s')]\rangle\langle\{\hat \Pi_{\mathcal{E}_b}(t),\mathcal{\hat E}_b(t')\}\rangle\langle[\hat\sigma(t'),\hat\sigma(s')]\rangle/2,
\end{align}
where the expectation values are taken in the undisturbed subsystems. We already omitted terms which will be proportional to $\langle\mathcal{\hat E}(t)\rangle=0$. The expectation value of the weak measurement output is given as
\begin{equation}
 \langle\mathcal{E}_a(t)\mathcal{E}_b(t)\rangle_w=\int\text{d}\mathcal{E}_a\text{d}\mathcal{E}_b\,p_{ab}(t)\mathcal{E}_a\mathcal{E}_b.
\end{equation} 
With $\int\text{d}\mathcal{E}\,\mathcal{E}\,\hat \Pi_{\mathcal{E}}(t)=\mathcal{\hat E}(t)$ and by introducing the response function $\chi(t,t')=-i\eta\theta(t-t')\langle [\mathcal{\hat E}(t),\mathcal{\hat E}(t')] \rangle$ and the noise function $S(t,t')=\frac{\eta}{2}\langle\{\mathcal{\hat E}(t),\mathcal{\hat E}(t')\}\rangle$ the field-field correlator can be expressed as
\begin{equation}
\langle\mathcal{E}(t)\mathcal{E}(t)\rangle_w=\frac{1}{2\hbar^2}\int\text{d}t'\int\text{d}s'\,\chi(t,t')\chi(t,s')\langle\{\hat\sigma(t'),\hat\sigma(s')\}\rangle
-\frac{2i}{\hbar^2}\int^{t}\text{d}t'\int^{t}\text{d}s'\,\theta(t'-s')S(t,t')\chi(t,s')\langle[\hat\sigma(t'),\hat\sigma(s')]\rangle,
\label{field-field-correlator}
\end{equation}
where we set $a=b$. In this case the cross-correlation differs from the measured quantity --- the intensity of the laser beam which corresponds to the self-correlation --- only by the intrinsic noise $\langle \mathcal{\hat E}^2(t)\rangle$. This is due to the scalar commutator of the electric field $\{[\mathcal{\hat E}^2(t),\mathcal{\hat E}(t')],\mathcal{\hat E}(s')\}=[\mathcal{\hat E}(t),\mathcal{\hat E}(t')]\{\mathcal{\hat E}(t),\mathcal{\hat E}(s')\}$. Since the intrinsic electric-field noise is independent of the measured system, it does not contribute to the absorption difference $A-A_0$ in which we are interested.

With the operator $\mathcal{\hat E}(t)=\frac{i}{\sqrt{2}}\left(\hat ae^{-i\omega t}-\hat a^{\dagger}e^{i\omega t}\right)$ we have the correlators
\begin{align}
[\mathcal{\hat E}(t),\mathcal{\hat E}(s)]=&-i\sin(\omega(t-s)),\nonumber\\
\{\mathcal{\hat E}(t),\mathcal{\hat E}(s)\}=&2\hat N\cos(\omega (t-s))-e^{i\omega(t+s)}(\hat a^{\dagger})^2-e^{-i\omega(t+s)}\hat a^2,
\end{align}
with $\hat N=\hat a^{\dagger}\hat a+\frac{1}{2}$. Note that the second term in Eq. (\ref{field-field-correlator}) is proportional to the photon number $N$ whereas the first is not. Therefore the system operators commutator $\langle[\hat\sigma(t'),\hat\sigma(s')]\rangle$ dominates the result for $N\gg1$.

\end{widetext}

\bibliography{coherence.bib}

\end{document}